# A quantum algorithm for evolving open quantum dynamics on quantum computing devices

Zixuan Hu[1,2], Rongxin Xia[1] and Sabre Kais*[1]

1. Department of Chemistry, Department of Physics, and Birck Nanotechnology Center, Purdue University, West Lafayette, IN 47907, United States
2. Qatar Environment and Energy Research Institute, College of Science and Engineering, HBKU, Doha, Qatar
   *Email: kais@purdue.edu

Designing quantum algorithms for simulating quantum systems has seen enormous progress, yet few studies have been done to develop quantum algorithms for open quantum dynamics despite its importance in modeling the system-environment interaction found in most realistic physical models. In this work we propose and demonstrate a general quantum algorithm to evolve open quantum dynamics on quantum computing devices. The Kraus operators governing the time evolution can be converted into unitary matrices with minimal dilation guaranteed by the Sz.-Nagy theorem. This allows the evolution of the initial state through unitary quantum gates, while using significantly less resource than required by the conventional Stinespring dilation. We demonstrate the algorithm on an amplitude damping channel using the IBM Qiskit quantum simulator and the IBM Q 5 Tenerife quantum device. The proposed algorithm does not require particular models of dynamics or decomposition of the quantum channel, and thus can be easily generalized to other open quantum dynamical models.

# I. Introduction

The time evolution of quantum systems is a century old subject that has been extensively studied for both fundamental and practical purposes. Open quantum dynamics is an important subfield of quantum physics that studies the time evolution of a system interacting with an environment[1]. Because the environment is usually too large to be treated exactly, in open quantum dynamics we often make approximations by averaging out the environment's effect on the system. The resultant time evolution of the system density matrix is non-unitary and often governed by a master equation. The idea of simulating quantum systems with quantum algorithms was first proposed by Feynman[2] and has received massive attention in recent years[3-17] for its promise of outperforming the best available classical algorithms. However relatively few studies[18-22] have been done to develop quantum algorithms for open quantum dynamics despite its importance. A key difficulty is the evolution of an open quantum system is often non-unitary, while quantum algorithms are mostly realized by unitary quantum gates. An early study[18] tackled this problem by including the environment in the quantum simulation process therefore making the evolution unitary. Focused on Markovian dynamics, their procedure required a reset on the environment for every time step, which can become expensive if the system becomes entangled with the environment or the evolution time is large. It is known that any non-unitary quantum operation can be made into a unitary one by the Stinespring dilation theorem[23]. However, due to the large increase of the dimension of the Hilbert space, the computational cost required by naïve application of Stinespring dilation to the quantum operation can be prohibitive for actual implementation on a quantum computing device. Perhaps for this reason, most algorithms developed (see e.g. Ref. [19,20]) so far rely on the knowledge of the decomposition of the quantum channel which may not be generally available for a large system without costly computational efforts. So far as we know a general quantum algorithm to simulate an arbitrary quantum channel for a general density matrix with minimal resource has not been proposed and demonstrated. In this study we propose and demonstrate such a quantum algorithm utilizing the Sz.-Nagy dilation theorem – being a variation of the Stinespring dilation theorem the Sz.-Nagy dilation requires much smaller dimension increase and can save significant computational resources. Without assuming any specific property of the quantum channel, we work with the most general form of the time evolution for a density matrix – the operator sum representation:

$$\rho(t) = \sum_k \rho_k(t) = \sum_k \mathbf{M}_k \rho \mathbf{M}_k^\dagger \tag{1}$$

where $\rho(t)$ is the system density matrix at time $t$, $\rho$ is the initial system density matrix, and $\mathbf{M}_k$'s are the Kraus operators that satisfy:

$$\sum_k \mathbf{M}_k^\dagger \mathbf{M}_k = \mathbf{I} \tag{2}$$

It is also common that the time evolution of $\rho(t)$ is described by a master equation:

$$\dot{\rho}(t) = \mathcal{L}\rho(t) \tag{3}$$

where the time derivative of $\rho(t)$ is given by the superoperator $\mathcal{L}$ applying to $\rho(t)$ itself. It is known that any master equation of the form of Eq. (3) can be converted into the form of Eq. (1) [24,25], affirming the generality of Eq. (1). An example of converting a widely used type of master equation – the Lindblad equation – into the operator sum representation has been provided in Ref. [26]. In this work we focus on the more general operator sum representation. Our method starts with an initial $\rho$ expressed by a sum of different pure quantum states weighted by the associated probabilities:

$$\rho = \sum_i p_i |\phi_i\rangle\langle\phi_i| \quad (4)$$

where $p_i$ is the probability of finding the state $|\phi_i\rangle$ in the mixture with the condition $\sum_i p_i = 1$. Note here the different $|\phi_i\rangle$'s are not necessarily orthogonal to each other, and Eq. (4) should be understood as a knowledge of the initial physical composition of the system that is easily available from system preparation, not as a diagonal decomposition of $\rho$ that requires considerable resource to compute. We show in the following that the quantum algorithm proposed can simulate the time evolution of $\rho(t)$. The outputs of the algorithm carry the full information of $\rho(t)$ that can be extracted by quantum tomography[27]. However, we remark that the most important physical information carried by $\rho(t)$ -- e.g. the populations of different quantum states and the expectation value of an observable $\langle \mathbf{O} \rangle = Tr(\mathbf{O}\rho(t))$ -- can be obtained without quantum tomography but by projection measurements instead, which greatly reduces the resources needed. This is realized by exploiting the positive-semidefiniteness of $\rho(t)$. We then estimate the gate complexity of our method to be significantly lower than the conventional Stinespring dilation and comparable to the classical method. Finally we demonstrate the application of the quantum algorithm to an amplitude damping channel with implementation on the IBM Qiskit[28] quantum simulator and the IBM Q 5 Tenerife[29] quantum device.

## II. Theory for the algorithm

In this section we present the quantum algorithm that evolves $\rho(t)$ with the initial $\rho$ given in the form $\rho = \sum_i p_i |\phi_i\rangle\langle\phi_i|$ in Eq.(4), which represents a knowledge of the initial physical composition of the system. Here we prepare each $|\phi_i\rangle$ as an input state $\mathbf{v}_i$ in a given basis and want to evolve:

$$|\phi_{ik}(t)\rangle = \mathbf{M}_k \mathbf{v}_i = (c_{ik1}, c_{ik2}, \ldots, c_{ikn})^T \quad (5)$$

First note that each Kraus operator $\mathbf{M}_k$ is a contraction such that it can be dilated into a unitary matrix. An operator $\mathbf{A}$ is a contraction if it shrinks or preserves the norm of any vector such that the operator norm $\|\mathbf{A}\| = \sup \frac{\|\mathbf{A}\mathbf{v}\|}{\|\mathbf{v}\|} \leq 1$. By Eq. (2) we have $\sum_k \mathbf{M}_k^\dagger \mathbf{M}_k = \mathbf{I}$, which implies $\mathbf{M}_k$ is a contraction (see the supplementary information (SI) for the proof). By the Sz.-Nagy dilation theorem[30,31], any contraction $\mathbf{A}$ of a Hilbert space $H$ has a corresponding unitary operator $\mathbf{U}_\mathbf{A}$ in a larger Hilbert space $K$ such that:

$$\mathbf{A}^n = \mathbf{P}_H \mathbf{U}_\mathbf{A}^n \mathbf{P}_H, \quad n \leq N \tag{6}$$

where $\mathbf{P}_H$ is the projection operator into space $H$. The physical meaning of Eq. (6) is that the effect of a contraction $\mathbf{A}$ applied up to $N$ times on a smaller space $H$ can be replicated by a unitary $\mathbf{U}_\mathbf{A}$ applied up to $N$ times on a larger space $K$, given the input vector lies entirely in $H$ and the output vector is projected into $H$. For the purpose of creating a quantum circuit the Sz.-Nagy dilation theorem allows us to simulate the effect of any non-unitary matrix by a unitary quantum gate, because every operator on a finite dimensional space is bounded and therefore can be made into a contraction which has a unitary dilation. The Sz.-Nagy theorem also guarantees the existence of a minimal dilation in the sense that the space $K$ has the smallest dimension to achieve Eq. (6). An example of a minimal unitary dilation of $\mathbf{A}$ with $N = 1$ is:

$$\mathbf{U}_\mathbf{A} = \begin{pmatrix} \mathbf{A} & \mathbf{D}_{\mathbf{A}^\dagger} \\ \mathbf{D}_\mathbf{A} & -\mathbf{A}^\dagger \end{pmatrix} \tag{7}$$

where $\mathbf{D}_\mathbf{A} = \sqrt{\mathbf{I} - \mathbf{A}^\dagger \mathbf{A}}$ is called the defect operator of $\mathbf{A}$. We can easily verify that $\mathbf{U}_\mathbf{A}$ is unitary and $\mathbf{A} = \mathbf{P}_H \mathbf{U}_\mathbf{A} \mathbf{P}_H$. However if we want to apply $\mathbf{A}^2 = \mathbf{P}_H \mathbf{U}_\mathbf{A}^2 \mathbf{P}_H$, or $\mathbf{BA} = \mathbf{P}_H \mathbf{U}_\mathbf{B} \mathbf{U}_\mathbf{A} \mathbf{P}_H$ where $\mathbf{U}_\mathbf{B}$ is a unitary dilation of $\mathbf{B}$, then we need minimal unitary dilations with $N = 2$:

$$\mathbf{U}_\mathbf{A} = \begin{pmatrix} \mathbf{A} & 0 & \mathbf{D}_{\mathbf{A}^\dagger} \\ \mathbf{D}_\mathbf{A} & 0 & -\mathbf{A}^\dagger \\ 0 & \mathbf{I} & 0 \end{pmatrix}, \quad \mathbf{U}_\mathbf{B} = \begin{pmatrix} \mathbf{B} & 0 & \mathbf{D}_{\mathbf{B}^\dagger} \\ \mathbf{D}_\mathbf{B} & 0 & -\mathbf{B}^\dagger \\ 0 & \mathbf{I} & 0 \end{pmatrix} \tag{8}$$

We see that the number $N$ in Eq. (6) is an important parameter that defines both the form and the applicability of the minimal unitary dilation. In the following we will refer to a minimal unitary dilation with a given $N$ an $N$-dilation[30]. A general rule is that to simulate the effect of $N$ contractions multiplying successively, we need to convert all of them to $N$-dilations. Going back to our original goal of simulating $\mathbf{M}_k \mathbf{v}_i$ with unitary gates we construct a 1-dilation $\mathbf{U}_\mathbf{A}$ of the form in Eq. (7) with $\mathbf{A} = \mathbf{M}_k$ and evolve:

$$|\phi_{ik}(t)\rangle = \mathbf{M}_k \mathbf{v}_i \xrightarrow{\text{unitary dilation}} \mathbf{U}_{\mathbf{M}_k} (\mathbf{v}_i^T, 0, ..., 0)^T \tag{9}$$

If $\mathbf{M}_k$ has the dimension $n$ by $n$, then the 1-dilation $\mathbf{U}_{\mathbf{M}_k}$ is $2n$ by $2n$. Now $\mathbf{U}_{\mathbf{M}_k}$ can be further decomposed into sequences of two-level unitary gates with a procedure illustrated in Ref. [3,32]. The two-level unitary gates can be used as elementary gates in an optical beamsplitter setup[32-35] and in the following we use the number of two-level unitary gates in the decomposition of a unitary gate to represent the gate complexity. Generally the number of two-level unitary gates needed to decompose a unitary gate is equal to the number of non-zero elements in the lower-triangular part of the gate[3,32]. For each $\mathbf{U}_{\mathbf{M}_k}$ of the form in Eq. (7) with $\mathbf{A} = \mathbf{M}_k$, we have $\frac{n^2 - n}{2}$ non-zero elements from $\mathbf{M}_k$, $n^2$ non-zero elements from $\mathbf{D}_{\mathbf{M}_k}$, $\frac{n^2 - n}{2}$ non-zero elements from $-\mathbf{M}_k^\dagger$, and the total gate count is $2n^2 - n$ for each $k$ and $i$. The classical complexity of evaluating $|\phi_{ik}(t)\rangle = \mathbf{M}_k \mathbf{v}_i$ is incidentally also $2n^2 - n$ for each $k$ and $i$, counting all the multiplications and additions needed for a matrix multiplying a vector. As a comparison, the conventional Stinespring dilation converts the whole quantum operation $\mathcal{E}(\rho) = \sum_k \mathbf{M}_k \rho \mathbf{M}_k^\dagger$ into a unitary operation $\mathbf{U}(\rho \otimes |e\rangle\langle e|) \mathbf{U}^\dagger$ by considering the tensor product space of the initial $\rho$ and an auxiliary environment[3]. The dimension of $\mathbf{U}$ is given by $n \cdot m$ where $m$ is the total number of Kraus operators $\mathbf{M}_k$ in the sum $\sum_k \mathbf{M}_k \rho \mathbf{M}_k^\dagger$. In the most general case $m = n^2$, $\mathbf{U}$ is $n^3$ by $n^3$ which requires $\frac{n^6 - n^3}{2}$ two-level unitary gates for the gate decomposition. We see that our procedure involving the Sz.-Nagy dilation offers a major simplification over the conventional Stinespring dilation: $2n^2 - n$ two-level $2n$ by $2n$ gates versus $\frac{n^6 - n^3}{2}$ two-level $n^3$ by $n^3$ gates. This is achieved by breaking the quantum operation $\mathcal{E}(\rho) = \sum_k \mathbf{M}_k \rho \mathbf{M}_k^\dagger$ into single $\mathbf{M}_k$ pieces and evolving them separately. Of course for $m = n^2$ we need $n^2$ circuits to evolve all the $\mathbf{M}_k$'s, but this can be done in parallel, thus the complexity of each single circuit is greatly reduced. Now to put the pieces back together, the full evolved density matrix can be assembled by $\rho(t) = \sum_{ik} p_i \cdot |\phi_{ik}(t)\rangle\langle\phi_{ik}(t)|$ and measured by quantum tomography. However, we remark that the most important physical information such as the populations of different states and the expectation value of an observable can be extracted from the output $|\phi_{ik}(t)\rangle$ not by quantum tomography, but by projection measurements instead, thus saving considerable computational costs. To get the populations of states in the current basis, note the diagonal vector of $|\phi_{ik}(t)\rangle\langle\phi_{ik}(t)|$ is:

$$diag\left(|\phi_{ik}(t)\rangle\langle\phi_{ik}(t)|\right) = \left(|c_{ik1}|^2, |c_{ik2}|^2, \ldots, |c_{ikn}|^2\right)^T \qquad (10)$$

Eq. (10) means we can obtain the diagonal element $|c_{ikj}|^2$ of $|\phi_{ik}(t)\rangle\langle\phi_{ik}(t)|$ by applying a projection measurement on the $j^{th}$ entry in the first $n$-dimensional subspace of $\mathbf{U}_{\mathbf{M}_k}\left(\mathbf{v}_i^T, 0, \ldots, 0\right)^T$. Using an optical setup such as in Ref. [35] the probability of measuring each entry in $\mathbf{U}_{\mathbf{M}_k}\left(\mathbf{v}_i^T, 0, ..., 0\right)^T$ can be efficiently obtained by recording the photon distribution at the output of the optical modes. Adding the results through $k$ and $i$ gives the diagonal elements of $\rho(t)$:

$$diag(\rho(t)) = \sum_{ik} p_i \cdot diag\left(|\phi_{ik}(t)\rangle\langle\phi_{ik}(t)|\right) \tag{11}$$

which gives the populations in the current basis. Although the off-diagonal elements of $\rho(t)$ cannot be directly obtained without quantum tomography, they are nonetheless carried by $|\phi_{ik}(t)\rangle$ and can become physically important. For example, if we want to get the populations in another basis, a unitary basis transformation can be applied to each $|\phi_{ik}(t)\rangle$ before measuring the diagonal elements:

$$\mathbf{T}|\phi_{ik}(t)\rangle = \mathbf{TM}_k\mathbf{v}_i \xrightarrow{\text{unitary dilation}} \begin{pmatrix} \mathbf{T} & 0 \\ 0 & \mathbf{I} \end{pmatrix}\mathbf{U}_{\mathbf{M}_k}\left(\mathbf{v}_i^T, 0, ..., 0\right)^T$$

$$diag(\mathbf{T}\rho(t)\mathbf{T}^\dagger) = \sum_{ik} p_i \cdot diag\left(\mathbf{T}|\phi_{ik}(t)\rangle\langle\phi_{ik}(t)|\mathbf{T}^\dagger\right) \tag{12}$$

The additional unitary $\mathbf{T}$ applied to $|\phi_{ik}(t)\rangle$ requires no dilation and increases the quantum gate count by $\frac{n^2-n}{2}$ (non-zero elements in the lower-triangular part of $\mathbf{T}$) to a total of $\frac{5n^2-3n}{2}$ for each $k$ and $i$. The classical complexity of evaluating $\mathbf{TM}_k\mathbf{v}_i$ is doubled from $2n^2-n$ to $4n^2-2n$ for each $k$ and $i$. Note here the quantum algorithm outperforms the classical one by taking advantage of the unitarity of $\mathbf{T}$.

Next to evaluate the expectation value of an observable $\langle\mathbf{O}\rangle = Tr(\mathbf{O}\rho(t))$, we recognize $\mathbf{O}\rho(t)$ is not always positive-semidefinite and hence additional processing must be done before the trace can be obtained from projection measurements on the output quantum state. The idea here is to see the operator norm is bounded by the Hilbert-Schmidt norm: $\|\mathbf{O}\| \leq \|\mathbf{O}\|_{HS}$, such that we define:

$$\tilde{\mathbf{O}} = \frac{\mathbf{O} + \mathbf{I}\|\mathbf{O}\|_{HS}}{2\|\mathbf{O}\|_{HS}} \tag{13}$$

where $\tilde{\mathbf{O}}$ is always a contraction and positive-semidefinite (see the SI for a detailed proof) so we can apply the Cholesky decomposition[36]: $\tilde{\mathbf{O}} = \mathbf{LL}^\dagger$, where $\mathbf{L}$ is a lower triangular matrix. Note we could have defined $\tilde{\mathbf{O}}$ with $\|\mathbf{O}\|$ in place of $\|\mathbf{O}\|_{HS}$ in Eq. (13), but $\|\mathbf{O}\|_{HS}$ requires fewer arithmetic steps to calculate. Now we have:

$$\langle\tilde{\mathbf{O}}\rangle = Tr(\tilde{\mathbf{O}}\rho(t)) = Tr(\mathbf{LL}^\dagger\rho(t)) = Tr(\mathbf{L}^\dagger\rho(t)\mathbf{L}) = \sum_k Tr(\mathbf{L}^\dagger\rho_k(t)\mathbf{L}) \tag{14}$$

where each $\mathbf{L}^\dagger \rho_k(t) \mathbf{L}$ is positive-semidefinite. $\mathbf{L}^\dagger$ is obviously a contraction because $\tilde{\mathbf{O}}$ is a contraction. Now evolve $\mathbf{L}^\dagger |\phi_{ik}(t)\rangle$ with two 2-dilations of the form in Eq. (8) with $\mathbf{A} = \mathbf{M}_k$ and $\mathbf{B} = \mathbf{L}^\dagger$:

$$\mathbf{L}^\dagger |\phi_{ik}(t)\rangle = \mathbf{L}^\dagger \mathbf{M}_k \mathbf{v}_i \xrightarrow{\text{unitary dilation}} \mathbf{U}_{\mathbf{L}^\dagger} \mathbf{U}_{\mathbf{M}_k} (\mathbf{v}_i^T, 0, ..., 0)^T \quad (15)$$

and we can evaluate $\langle \tilde{\mathbf{O}} \rangle$ by:

$$\langle \tilde{\mathbf{O}} \rangle = Tr(\tilde{\mathbf{O}} \rho(t)) = \sum_{i,k} Tr\left( p_i \cdot \mathbf{L}^\dagger |\phi_{ik}(t)\rangle \langle \phi_{ik}(t)| \mathbf{L} \right) \quad (16)$$

where the trace of $\mathbf{L}^\dagger |\phi_{ik}(t)\rangle \langle \phi_{ik}(t)| \mathbf{L}$ can be obtained by projection measurements in a way similar to what has been explained for Eq. (10). The difference here is that we do not need to measure individual diagonal elements and sum them together for the trace, but instead can measure the total probability of projecting into the first $n$-dimensional space of $\mathbf{U}_{\mathbf{L}^\dagger} \mathbf{U}_{\mathbf{M}_k} (\mathbf{v}_i^T, 0, ..., 0)^T$. This can potentially save measurement costs by reducing the number of inquiries needed on the output vector. Finally $\langle \mathbf{O} \rangle$ is calculated with:

$$\begin{aligned} \langle \mathbf{O} \rangle &= Tr(\mathbf{O} \rho(t)) \\ &= Tr\left( \left( 2\|\mathbf{O}\|_{HS} \tilde{\mathbf{O}} - \mathbf{I} \|\mathbf{O}\|_{HS} \right) \rho(t) \right) \\ &= 2\|\mathbf{O}\|_{HS} Tr(\tilde{\mathbf{O}} \rho(t)) - \|\mathbf{O}\|_{HS} \end{aligned} \quad (17)$$

where we have successfully obtained $\langle \mathbf{O} \rangle$ for the original observable.

The $\mathbf{L}^\dagger$ gate ($\mathbf{L}^\dagger$ is upper triangular requiring reduced number of two-level unitaries for the decomposition) plus the additional level of dilation for $\mathbf{M}_k$ increases the quantum gate count to $\frac{5n^2 + 3n}{2}$ for each $k$ and $i$. A classical overhead (independent from the $k$ and $i$ counts) cost of $2n^2 - 1$ for $\|\mathbf{O}\|_{HS}$ and $\frac{n^3}{3}$ for the Cholesky decomposition[36] should also be counted towards the total cost of the quantum algorithm. On the other hand the classical complexity of evaluating $\mathbf{L}^\dagger \mathbf{M}_k \mathbf{v}_i$ is $3n^2 - n$ for each $k$ and $i$ (taking into account that $\mathbf{L}^\dagger$ is upper triangular), plus an overhead of $\frac{n^3}{3} + 2n^2 - 1$.

### III. Application to the amplitude damping channel

In this section we use a quantum channel model to demonstrate the application of the proposed method. The amplitude-damping channel models the spontaneous emission of a 2-level atom. The Lindblad master equation for this model is:

$$\dot{\rho}(t) = \gamma\left(\sigma^+ \rho(t) \sigma^- - \frac{1}{2}\{\sigma^- \sigma^+, \rho(t)\}\right) \quad (18)$$

where $\gamma$ is the spontaneous emission rate, $\sigma^+ = |0\rangle\langle 1|$ is the Pauli raising operator that mediates the transition from the excited state to the ground state, and $\sigma^- = (\sigma^+)^\dagger$. In the operator sum representation:

$$\rho(t) = \mathbf{M}_0 \rho \mathbf{M}_0^\dagger + \mathbf{M}_1 \rho \mathbf{M}_1^\dagger \quad (19)$$

$$\mathbf{M}_0 = \frac{1+\sqrt{e^{-\gamma t}}}{2}\mathbf{I} + \frac{1-\sqrt{e^{-\gamma t}}}{2}\sigma_z = \begin{pmatrix} 1 & 0 \\ 0 & \sqrt{e^{-\gamma t}} \end{pmatrix} \quad \mathbf{M}_1 = \sqrt{1-e^{-\gamma t}}\sigma^+ = \begin{pmatrix} 0 & \sqrt{1-e^{-\gamma t}} \\ 0 & 0 \end{pmatrix}$$

To calculate the populations in the initial basis $\{|0\rangle, |1\rangle\}$ we can construct $\mathbf{U}_{\mathbf{M}_0}$ and $\mathbf{U}_{\mathbf{M}_1}$ of the 1-dilation form in Eq. (7) with $\mathbf{A} = \mathbf{M}_k$ and $\mathbf{D}_{\mathbf{M}_k} = \sqrt{\mathbf{I} - \mathbf{M}_k^\dagger \mathbf{M}_k}$:

$$\mathbf{D}_{\mathbf{M}_0} = \begin{pmatrix} 0 & 0 \\ 0 & \sqrt{1-e^{-\gamma t}} \end{pmatrix}, \quad \mathbf{D}_{\mathbf{M}_1} = \begin{pmatrix} 1 & 0 \\ 0 & \sqrt{e^{-\gamma t}} \end{pmatrix} \quad (20)$$

Using an arbitrary initial $\rho = \frac{1}{4}\begin{pmatrix} 1 & 1 \\ 1 & 3 \end{pmatrix}$ and assuming the physical composition $\rho = \frac{1}{2}(|1\rangle\langle 1| + |+\rangle\langle +|)$ is known, the input states are:

$$|\phi_1\rangle = |1\rangle \rightarrow \mathbf{v}_1 = \left(0, 1, \overbrace{0, ..., 0}^{m}\right)^T \quad |\phi_2\rangle = |+\rangle \rightarrow \mathbf{v}_2 = \frac{1}{\sqrt{2}}\left(1, 1, \overbrace{0, ..., 0}^{m}\right)^T \quad (21)$$

where $m = 2$ matching the size of the vectors with the dilation $\mathbf{U}_{\mathbf{M}_k}$. We set $\gamma = 1.52 \times 10^9 \text{s}^{-1}$ (typical nanosecond lifetime), numerically calculate $\mathbf{U}_{\mathbf{M}_k} \mathbf{v}_i$ from $t = 0$ to $t = 1000$ ps with a time step of 10 picosecond, and obtain the populations of the ground and excited states from the first two entries of $\mathbf{U}_{\mathbf{M}_k} \mathbf{v}_i$. The results are shown as the smooth lines in Figure 1.

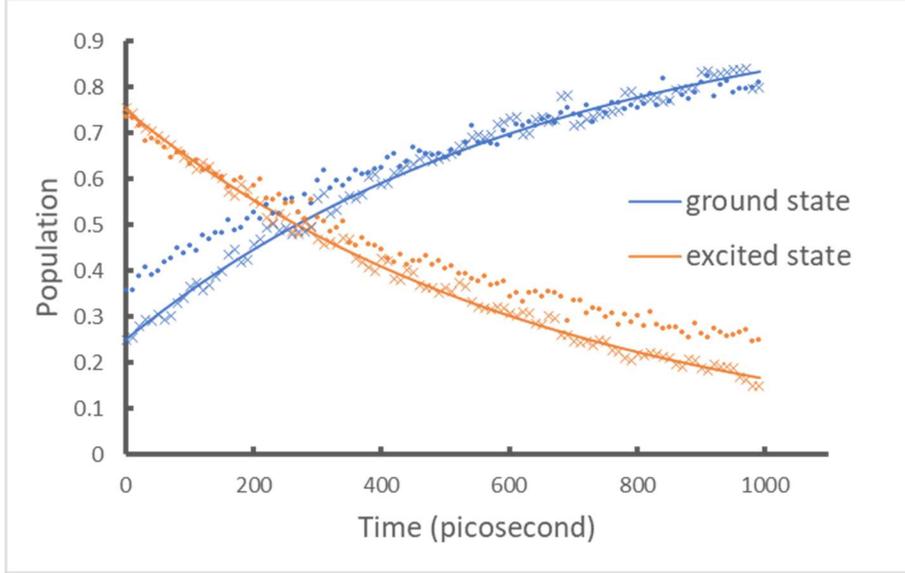

Figure 1. Showing the populations of the ground and excited states for the amplitude damping model. The smooth lines are obtained by classical numerical calculations of the output vectors. These agree exactly with analytic results and are used as benchmarks. The crosses are obtained by the IBM Qiskit simulator. The dots are obtained by the IBM Q 5 Tenerife device. The quantum circuits include 2 qubits and on average 13 elementary gates (see Figure 4 for an example and the SI for all the circuits).

To calculate the populations in another basis $\{|+\rangle, |-\rangle\}$ where $|\pm\rangle = \frac{1}{\sqrt{2}}(|0\rangle \pm |1\rangle)$, we need the transformation matrix $\mathbf{T} = \frac{1}{\sqrt{2}}\begin{pmatrix} 1 & 1 \\ 1 & -1 \end{pmatrix}$. We numerically calculate $\begin{pmatrix} \mathbf{T} & 0 \\ 0 & \mathbf{I} \end{pmatrix} \mathbf{U}_{\mathbf{M}_k} \mathbf{v}_i$ and obtain the populations of the $|\pm\rangle$ states shown as the smooth lines in Figure 2.

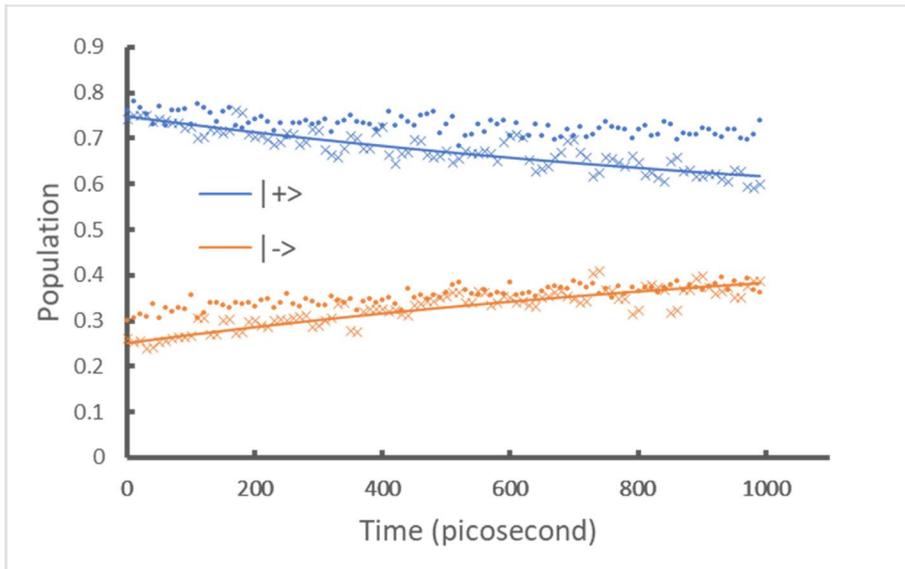

Figure 2. Showing the populations of the $|+\rangle$ and $|-\rangle$ states for the amplitude damping model. The smooth lines are obtained by classical numerical calculations of the output vectors. These agree exactly with analytic results and are used as benchmarks. The crosses are obtained by the IBM Qiskit simulator. The dots are obtained by the IBM Q 5 Tenerife device. The quantum circuits include 2 qubits and on average 30 elementary gates (see the SI for the circuits).

Now we evaluate the expectation value of an observable $\langle \mathbf{O} \rangle$ for $\mathbf{O} = \begin{pmatrix} -2 & 0.5 \\ 0.5 & 1 \end{pmatrix}$ as an example. With $\|\mathbf{O}\|_{HS} = \frac{\sqrt{22}}{2} \approx 2.35$, we define $\tilde{\mathbf{O}} = \frac{\mathbf{O} + \mathbf{I}\|\mathbf{O}\|_{HS}}{2\|\mathbf{O}\|_{HS}} \approx \begin{pmatrix} 0.0740 & 0.107 \\ 0.107 & 0.713 \end{pmatrix}$ and find $\mathbf{L} \approx \begin{pmatrix} 0.271 & 0 \\ 0.393 & 0.748 \end{pmatrix}$ through Cholesky decomposition $\tilde{\mathbf{O}} = \mathbf{L}\mathbf{L}^{\dagger}$. Next we construct $\mathbf{U}_{\mathbf{L}^{\dagger}}$ and $\mathbf{U}_{\mathbf{M}_k}$ of the 2-dilation form in Eq. (8) and apply them to the initial state $\mathbf{v}_i$ in the form of Eq. (21) with $m = 4$. Numerically calculating the output vector will give us $\langle \mathbf{O} \rangle$ by Eq. (16) and (17). The results are shown as the smooth line in Figure 3:

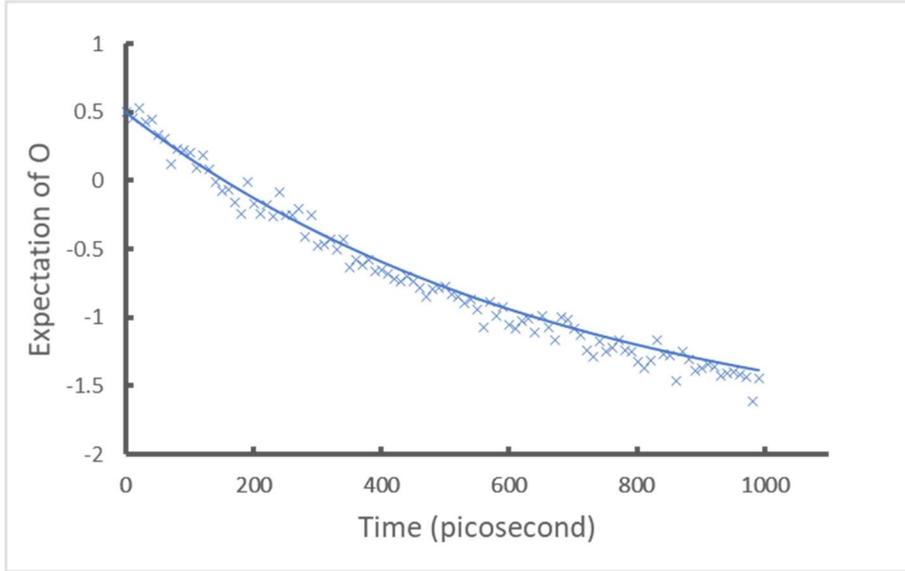

Figure 3. Showing the expectation values $\langle \mathbf{O} \rangle$. The smooth line is obtained by classical numerical calculations of the output vectors. This agrees exactly with analytic results and is used as a benchmark. The crosses are obtained by the IBM Qiskit simulator. The quantum circuits include 3 qubits and on average 184 elementary gates (see the SI for the circuits). Due to the large number of gates required the quantum device is not used for these results.

The overall simplicity of the unitary dilation gates used in our algorithm allows us to further demonstrate it using the IBM Qiskit[28] quantum simulator and the IBM Q 5 Tenerife[29] quantum device. So far we have been using the 2-level unitaries obtained from the $\mathbf{U}_{\mathbf{M}_0}$ and $\mathbf{U}_{\mathbf{M}_1}$ gates as elementary gates in our complexity evaluation. This is possible if we use an optical platform for quantum computation involving beamsplitters[32,34]. To implement the algorithm on a conventional quantum platform such as the IBM simulator and devices, we further decompose the 2-level unitaries into elementary 1-qubit and 2-qubit gates. The constructed circuits require 2 qubits and on average 13 elementary gates for the basic $\mathbf{U}_{\mathbf{M}_k \mathbf{v}_i}$ evolution and 30 elementary gates for the basis transformation $\begin{pmatrix} \mathbf{T} & 0 \\ 0 & \mathbf{I} \end{pmatrix} \mathbf{U}_{\mathbf{M}_k \mathbf{v}_i}$. An example of the circuit for $\mathbf{U}_{\mathbf{M}_0 \mathbf{v}_1}$ is shown in Figure 4, and a full list of the circuits can be found in the SI.

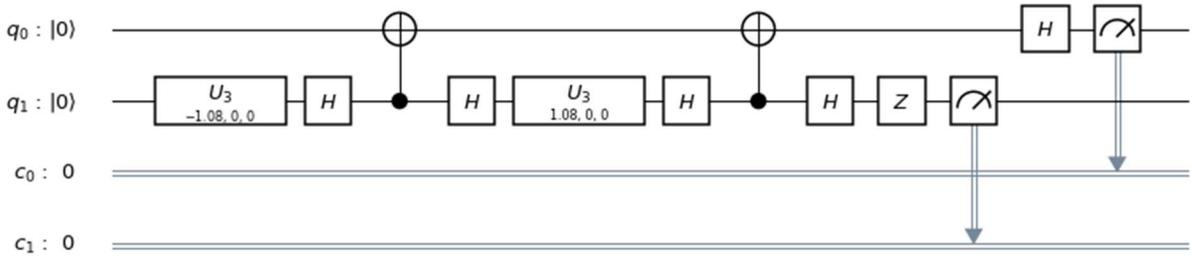

Figure 4. Showing the quantum circuit for $\mathbf{U}_{\mathbf{M}_0 \mathbf{v}_1}$ as used in the IBM simulator and quantum device. For a full list of the quantum circuits please see the SI.

We implement these circuits on both the simulator and the quantum devices and show the results as the crosses (simulator) and dots (device) in Figure 1 and Figure 2. To obtain $\langle \tilde{\mathbf{O}} \rangle$ we construct a circuit of 3 qubits with an average of 182 elementary gates for the $\mathbf{U}_{\mathbf{L}^\dagger} \mathbf{U}_{\mathbf{M}_k \mathbf{v}_i}$ operation (see the SI). The drastic increase in the number of gates is due to the increased size of the $\mathbf{U}_{\mathbf{L}^\dagger}$ and $\mathbf{U}_{\mathbf{M}_k}$ matrices. The decomposition of 2-level unitaries into 1-qubit and 2-qubit elementary gates grows rapidly with the overall size of the 2-level unitaries because of the complexity to implement multi-qubit control gates. The large number of gates for this circuit prevents us from running it on the quantum device so that only the simulator results are shown as the crosses in Figure 3. On the other hand the 2-level unitaries can be more easily implemented with a multiport photonic device as demonstrated in Refs. [34,35]. We will seek to demonstrate our methods on such a preferred device in a future study.

In Figure 1, Figure 2 and Figure 3, the numerical results (smooth lines) agree exactly with analytic results and are used as benchmarks for the simulator and device results. The simulator results (crosses) fit the benchmark well while including the probabilistic error from the projection measurements. The quantum device results (dots) fit the benchmark reasonably well considering all experimental errors from gate fault, qubit decoherence, and measurement. These results demonstrate the ability of the proposed algorithm to simulate the time evolution of a density matrix and evaluate physical observables from the outputs.

## IV. Discussion and conclusion

In this work we have presented a quantum algorithm for evolving the density matrix $\rho(t) = \sum_k \mathbf{M}_k \rho \mathbf{M}_k^\dagger$ and extracting physical information from the output. The method takes each physical composition $|\phi_i\rangle$ of $\rho = \sum_i p_i |\phi_i\rangle\langle\phi_i|$ as the input and evolves it through minimal Sz.-Nagy dilations of the Kraus operators $\mathbf{M}_k$. The input vector has the base length $n$ (plus additional zeros matching the dimension of the evolution matrices) and various realizations of the time evolution have the quantum gate count of $O(n^2)$ for each $k$ and $i$, which is a significant improvement over the $O(n^6)$ scaling of the conventional Stinespring dilation. The requirement of knowing the initial physical composition as a probability-weighted mixture of not necessarily orthogonal $|\phi_i\rangle$'s should be easy to satisfy from system preparation. If indeed the initial physical composition is unknown and we have to work with the most general matrix form of the density matrix, then the method needs to be modified. As details shown in the SI, the modified method uses the flattened vector of the initial density matrix as the input and requires $O(n^3)$ for various realizations of the time evolution. Both methods can be easily generalized to other open quantum dynamical models because the procedures involved are essentially the same – only the Kraus operators $\mathbf{M}_k$'s for the operator sum representation are different for different models. The generality of our methods -- in the sense of not requiring particular dynamical models or costly decompositions of the density matrix and the quantum channel -- opens up the possibility of simulating more interesting systems such as decohering qubits or excitonic structures interacting with multiple baths[37,38] – the latter helps to understand natural light harvesting complexes and exploit quantum coherence effect to improve light harvesting efficiency in artificial photocells[37,39]. Finally we have demonstrated the implementation of the algorithm on the IBM Qiskit simulator and the IBM Q 5 Tenerife device. Although the gate complexity is larger than calculated with the preferred optical setup, the results show reasonable agreements with the analytic benchmarks considering gate fault, qubit decoherence, and measurement error. In future studies we will seek to demonstrate our quantum algorithms on the preferred photonic devices that can implement 2-level unitaries as elementary gates.

## V. Acknowledgement

We acknowledge the funding from the U.S. Department of Energy, Office of Basic Energy Sciences under award number DE-SC0019215, the Qatar National Research Fund exceptional Grant NPRPX-107-1-027 and the National Science Foundation under award number 1839191-ECCS.

## Supplementary information: A quantum algorithm for evolving open quantum dynamics on quantum computing devices

Zixuan Hu[1,2], Rongxin Xia[1] and Sabre Kais*[1]

3. Department of Chemistry, Department of Physics, and Birck Nanotechnology Center, Purdue University, West Lafayette, IN 47907, United States
4. Qatar Environment and Energy Research Institute, College of Science and Engineering, HBKU, Doha, Qatar
   *Email:* kais@purdue.edu

This supplementary document supports the discussion in the main text by providing technical details. Section 1 provides the proof that each Kraus operator $\mathbf{M}_k$ is a contraction. Section 2 proves that $\tilde{\mathbf{O}}$ is a contraction and positive-semidefinite. Section 3 presents the modified algorithm for the initial state given in a general matrix form. Section 4 lists the quantum circuits used on the IBM Qiskit simulator and the IBM Q 5 Tenerife device.

1. **Proof that each Kraus operator $\mathbf{M}_k$ is a contraction**

As defined in the main text, an operator $\mathbf{A}$ is a contraction if it shrinks or preserves the norm of any vector such that the operator norm $\|\mathbf{A}\| = \sup \frac{\|\mathbf{A}\mathbf{v}\|}{\|\mathbf{v}\|} \leq 1$. We have $\sum_k \mathbf{M}_k^\dagger \mathbf{M}_k = \mathbf{I}$, suppose an arbitrary one $\mathbf{M}_1$ is not a contraction, then for $\mathbf{M}_1$ there exists a vector $\mathbf{v}_1$ such that

$$\frac{\|\mathbf{M}_1\mathbf{v}_1\|}{\|\mathbf{v}_1\|} = \frac{\sqrt{\mathbf{v}_1^\dagger \mathbf{M}_1^\dagger \mathbf{M}_1 \mathbf{v}_1}}{\|\mathbf{v}_1\|} > 1, \text{ then } \frac{\sqrt{\mathbf{v}_1^\dagger \sum_k \mathbf{M}_k^\dagger \mathbf{M}_k \mathbf{v}_1}}{\|\mathbf{v}_1\|} > \frac{\sqrt{\mathbf{v}_1^\dagger \mathbf{M}_1^\dagger \mathbf{M}_1 \mathbf{v}_1}}{\|\mathbf{v}_1\|} > 1 \text{ because each } \mathbf{M}_k^\dagger \mathbf{M}_k \text{ is}$$

positive-semidefinite with $\mathbf{v}_1^\dagger \mathbf{M}_k^\dagger \mathbf{M}_k \mathbf{v}_1 \geq 0$. However because $\sum_k \mathbf{M}_k^\dagger \mathbf{M}_k = \mathbf{I}$ we also have

$$\frac{\sqrt{\mathbf{v}_1^\dagger \sum_k \mathbf{M}_k^\dagger \mathbf{M}_k \mathbf{v}_1}}{\|\mathbf{v}_1\|} = \frac{\sqrt{\mathbf{v}_1^\dagger \mathbf{v}_1}}{\|\mathbf{v}_1\|} = 1.$$ Hence by contradiction each $\mathbf{M}_k$ is a contraction.

2. **Proof that $\tilde{\mathbf{O}} = \dfrac{\mathbf{O} + \mathbf{I}\|\mathbf{O}\|_{HS}}{2\|\mathbf{O}\|_{HS}}$ is a contraction and positive-semidefinite**

To prove $\tilde{\mathbf{O}} = \dfrac{\mathbf{O} + \mathbf{I}\|\mathbf{O}\|_{HS}}{2\|\mathbf{O}\|_{HS}}$ is a contraction:

$$\|\tilde{\mathbf{O}}\| = \left\|\frac{\mathbf{O} + \mathbf{I}\|\mathbf{O}\|_{HS}}{2\|\mathbf{O}\|_{HS}}\right\| \leq \frac{\|\mathbf{O}\| + \|\mathbf{I}\|\|\mathbf{O}\|_{HS}}{2\|\mathbf{O}\|_{HS}} \leq \frac{\|\mathbf{O}\|_{HS} + \|\mathbf{O}\|_{HS}}{2\|\mathbf{O}\|_{HS}} = 1 \tag{22}$$

Where we have used the triangle inequality of the operator norm and the fact that $\|\mathbf{O}\| \leq \|\mathbf{O}\|_{HS}$. Note we could have made $\tilde{\mathbf{O}} = \dfrac{\mathbf{O} + \mathbf{I}\|\mathbf{O}\|}{2\|\mathbf{O}\|}$ but the operator norm $\|\mathbf{O}\|$ is more difficult to calculate than the Hilbert-Schmidt norm $\|\mathbf{O}\|_{HS}$. To prove $\tilde{\mathbf{O}} = \dfrac{\mathbf{O} + \mathbf{I}\|\mathbf{O}\|_{HS}}{2\|\mathbf{O}\|_{HS}}$ is positive-semidefinite, let $\lambda_{min}$ be the smallest eigenvalue of $\mathbf{O}$, then $|\lambda_{min}| \leq \|\mathbf{O}\| \leq \|\mathbf{O}\|_{HS}$ and $\mathbf{v}^\dagger \mathbf{O} \mathbf{v} \geq \lambda_{min} \|\mathbf{v}\|^2$ for any $\mathbf{v}$. Now we have for any $\mathbf{v}$:

$$\mathbf{v}^\dagger \tilde{\mathbf{O}} \mathbf{v} = \frac{\mathbf{v}^\dagger \mathbf{O} \mathbf{v} + \|\mathbf{O}\|_{HS}\|\mathbf{v}\|^2}{2\|\mathbf{O}\|_{HS}} \geq \frac{(\lambda_{min} + \|\mathbf{O}\|_{HS})\|\mathbf{v}\|^2}{2\|\mathbf{O}\|_{HS}} \geq 0 \tag{23}$$

Therefore $\tilde{\mathbf{O}}$ is indeed positive-semidefinite.

3. **Modified algorithm for the initial state in a general matrix form**

In this section we present the quantum algorithm that evolves $\rho(t)$ with the initial $\rho$ given in a general matrix form:

$$\rho = \begin{pmatrix} \rho_{11} & \cdots & \rho_{1n} \\ \vdots & \ddots & \vdots \\ \rho_{n1} & \cdots & \rho_{nn} \end{pmatrix} \tag{24}$$

We first flatten $\rho$ into a vector form:

$$\rho \to \mathbf{v}_\rho = (\rho_{11}, \ldots, \rho_{1n}, \rho_{21}, \ldots, \rho_{2n}, \ldots \ldots, \rho_{n1}, \ldots, \rho_{nn})^T \tag{25}$$

for which the norm of $\mathbf{v}_\rho$ is given by:

$$\|\mathbf{v}_\rho\| = \sqrt{\sum_{ij} |\rho_{ij}|^2} = \|\rho\|_{HS} = \sqrt{Tr(\rho^2)} \leq 1 \tag{26}$$

where $\|\rho\|_{HS}$ is the Hilbert-Schmidt norm of the density matrix $\rho$. Eq. (26) connects the norm of $\mathbf{v}_\rho$ to the purity $Tr(\rho^2)$ which measures how much a mixture is $\rho$.

Now for each $k$ in $\sum_k \mathbf{M}_k \rho \mathbf{M}_k^\dagger$ the $\mathbf{M}_k$ multiplying from the left is converted into $\mathcal{M}_k = \mathbf{M}_k \otimes \mathbf{I}$, and the $\mathbf{M}_k^\dagger$ multiplying from the right is converted into $\mathcal{N}_k = \mathbf{I} \otimes \overline{\mathbf{M}}_k$ where $\otimes$ stands for the Kronecker product, and the bar over $\mathbf{M}_k$ stands for complex conjugation. It is easy to verify that:

$$\mathbf{M}_k \rho \mathbf{M}_k^\dagger \xleftrightarrow{equivalent} \mathcal{N}_k \mathcal{M}_k \mathbf{v}_\rho \tag{27}$$

The core idea of this quantum algorithm is to represent $\mathbf{v}_\rho$ with a quantum state and then simulate the effects of $\mathcal{N}_k \mathcal{M}_k$ with quantum gates. Firstly $\mathbf{v}_\rho$ can be normalized and represented by an initial quantum state. In the main text we have proven each $\mathbf{M}_k$ is a contraction. It follows immediately that $\mathcal{M}_k = \mathbf{M}_k \otimes \mathbf{I}$ and $\mathcal{N}_k = \mathbf{I} \otimes \overline{\mathbf{M}}_k$ are also contractions by the norm property of the Kronecker product. To simulate $\mathcal{N}_k \mathcal{M}_k \mathbf{v}_\rho$ with unitary gates, we need two 2-dilations as in Eq. (8) in the main text by setting $\mathbf{A} = \mathcal{M}_k$ and $\mathbf{B} = \mathcal{N}_k$:

$$\mathcal{N}_k \mathcal{M}_k \mathbf{v}_\rho \xrightarrow{\text{unitary dilation}} \mathbf{U}_{\mathcal{N}_k} \mathbf{U}_{\mathcal{M}_k} (\mathbf{v}_\rho^T, 0, \ldots, 0)^T \tag{28}$$

If $\mathbf{M}_k$ has the dimension $n$ by $n$, then $\mathcal{M}_k$ and $\mathcal{N}_k$ are $n^2$ by $n^2$, and the 2-dilations $\mathbf{U}_{\mathcal{M}_k}$ and $\mathbf{U}_{\mathcal{N}_k}$ are $3n^2$ by $3n^2$. Now keeping with the main text we further decompose $\mathbf{U}_{\mathcal{M}_k}$ and $\mathbf{U}_{\mathcal{N}_k}$ into sequences of two-level unitary gates and count them towards the gate complexity. For $\mathbf{U}_{\mathcal{M}_k}$ the lower-triangular part contains $\frac{n^2-n}{2}\cdot n$ non-zero elements from $\mathscr{M}_k$, $n^3$ non-zero elements from $\mathbf{D}_{\mathscr{M}_k}$, $n^2$ non-zero elements from $\mathbf{I}$, and the total count is $\frac{3n^3}{2}+\frac{n^2}{2}$. This count is the same for $\mathbf{U}_{\mathscr{N}_k}$, thus the total gate complexity to realize $\mathscr{N}_k\mathscr{M}_k\mathbf{v}_\rho$ is $3n^3+n^2$ for each $k$. Note the classical complexity to realize $\mathbf{M}_k\rho\mathbf{M}_k^\dagger$ by two matrix multiplications is $4n^3-2n^2$ (using the naïve algorithm counting the number of multiplications and additions) for each $k$, which is of the same order of our proposed quantum algorithm, with a minor difference in the leading coefficient.

Now the density matrix has been evolved, we proceed to extract physical information from the output $\mathbf{v}_k(t)=\mathscr{N}_k\mathscr{M}_k\mathbf{v}_\rho$. Firstly the diagonal elements of each $\rho_k(t)=\mathbf{M}_k\rho\mathbf{M}_k^\dagger$ are always non-negative because $\mathbf{M}_k\rho\mathbf{M}_k^\dagger$ is positive-semidefinite. This implies that the diagonal elements of $\rho_k(t)$ can be obtained by applying a projection measurement on corresponding entries in $\mathbf{v}_k(t)$. Using an optical setup such as in Ref. [1] the probability of measuring each entry in $\mathbf{v}_k(t)$ can be efficiently obtained by recording the photon distribution at the output of the optical modes. Adding the diagonal elements of $\rho_k(t)$ over $k$ gives us the diagonal elements of the final $\rho(t)=\sum_k\mathbf{M}_k\rho\mathbf{M}_k^\dagger$ which are the populations of the final system state in the basis currently in use. Although the off-diagonal elements of $\rho_k(t)$ cannot be directly obtained without quantum tomography, they are nonetheless carried by $\mathbf{v}_k(t)$ and can become physically important. For example, if we want to obtain the populations of the final system in another basis, we can carry out a basis transformation $\rho_k(t)\to\mathbf{T}\rho_k(t)\mathbf{T}^\dagger$, or correspondingly $\mathbf{v}_k(t)\to(\mathbf{I}\otimes\overline{\mathbf{T}})(\mathbf{T}\otimes\mathbf{I})\mathbf{v}_k(t)$, where the off-diagonal elements are required. Here $\mathbf{T}$ is unitary such that $\mathbf{T}\otimes\mathbf{I}$ and $\mathbf{I}\otimes\overline{\mathbf{T}}$ are unitary, and therefore no dilations are needed. For the additional $\mathbf{T}\otimes\mathbf{I}$ and $\mathbf{I}\otimes\overline{\mathbf{T}}$ gates, the quantum gate count is increased by $n^3-n^2$ to a total of $4n^3$ for each $k$, while the classical complexity adds an overhead cost of $4n^3-2n^2$ for $\rho(t)\to\mathbf{T}\rho(t)\mathbf{T}^\dagger$ (independent from the $k$ count) to the original $4n^3-2n^2$ for each $k$. We remark that when the total number of $\mathbf{M}_k$ operators is small, the quantum algorithm outperforms the classical one by taking advantage of the unitarity of $\mathbf{T}$. The off-diagonal elements of $\rho_k(t)$ are also important if we want to calculate the expectation value of an observable: $\langle\mathbf{O}\rangle=Tr(\mathbf{O}\rho(t))=\sum_k Tr(\mathbf{O}\rho_k(t))$. Here we use the same procedure as in the main text to define an operator $\tilde{\mathbf{O}}=\frac{\mathbf{O}+\mathbf{I}\|\mathbf{O}\|_{HS}}{2\|\mathbf{O}\|_{HS}}$, which is both a contraction and positive-semidefinite. Now by Cholesky decomposition we have $\tilde{\mathbf{O}}=\mathbf{L}\mathbf{L}^\dagger$ and:

$$\langle \tilde{\mathbf{O}} \rangle = Tr(\tilde{\mathbf{O}}\rho(t)) = Tr(\mathbf{L}\mathbf{L}^\dagger \rho(t)) = Tr(\mathbf{L}^\dagger \rho(t)\mathbf{L}) = \sum_k Tr(\mathbf{L}^\dagger \rho_k(t)\mathbf{L}) \quad (29)$$

where each $\mathbf{L}^\dagger \rho_k(t)\mathbf{L}$ is positive-semidefinite and its diagonal elements can be obtained by projection measurements. $\mathbf{L}^\dagger$ is obviously a contraction because $\tilde{\mathbf{O}}$ is a contraction. To realize

$$\mathbf{L}^\dagger \rho_k(t)\mathbf{L} \xleftrightarrow{equivalent} (\mathbf{I} \otimes \bar{\mathbf{L}}^\dagger)(\mathbf{L}^\dagger \otimes \mathbf{I})\mathcal{N}_k \mathcal{M}_k \mathbf{v}_\rho \quad (30)$$

we need all four matrices to be in the 4-dilation form:

$$\mathbf{U_A} = \begin{pmatrix} \mathbf{A} & 0 & 0 & 0 & \mathbf{D}_{\mathbf{A}^\dagger} \\ \mathbf{D_A} & 0 & 0 & 0 & -\mathbf{A}^\dagger \\ 0 & \mathbf{I} & 0 & 0 & 0 \\ 0 & 0 & \mathbf{I} & 0 & 0 \\ 0 & 0 & 0 & \mathbf{I} & 0 \end{pmatrix} \quad (31)$$

where $\mathbf{A} = (\mathbf{L}^\dagger \otimes \mathbf{I})$, $(\mathbf{I} \otimes \bar{\mathbf{L}}^\dagger)$, $\mathcal{N}_k$, or $\mathcal{M}_k$:

$$(\mathbf{I} \otimes \bar{\mathbf{L}}^\dagger)(\mathbf{L}^\dagger \otimes \mathbf{I})\mathcal{N}_k \mathcal{M}_k \mathbf{v}_\rho \xrightarrow{unitary\ dilation} \mathbf{U}_{(\mathbf{I} \otimes \bar{\mathbf{L}}^\dagger)} \mathbf{U}_{(\mathbf{L}^\dagger \otimes \mathbf{I})} \mathbf{U}_{\mathcal{N}_k} \mathbf{U}_{\mathcal{M}_k} (\mathbf{v}_\rho^T, 0, ..., 0)^T \quad (32)$$

Now after $\mathbf{L}^\dagger \rho_k(t)\mathbf{L}$ has been obtained $Tr(\tilde{\mathbf{O}}\rho(t))$ can be calculated, then we have:

$$\begin{aligned} \langle \mathbf{O} \rangle &= Tr(\mathbf{O}\rho(t)) \\ &= Tr((2\|\mathbf{O}\|_{HS}\tilde{\mathbf{O}} - \mathbf{I}\|\mathbf{O}\|_{HS})\rho(t)) \\ &= 2\|\mathbf{O}\|_{HS} Tr(\tilde{\mathbf{O}}\rho(t)) - \|\mathbf{O}\|_{HS} \end{aligned} \quad (33)$$

where we have successfully obtained $\langle \mathbf{O} \rangle$ for the original observable. The $(\mathbf{L}^\dagger \otimes \mathbf{I})$ and $(\mathbf{I} \otimes \bar{\mathbf{L}}^\dagger)$ gates ( $\mathbf{L}^\dagger$ is upper triangular requiring reduced number of two-level unitaries for the decomposition) plus the additional two levels of dilation for $\mathcal{M}_k$ and $\mathcal{N}_k$ increase the quantum gate count to $5n^3 + 11n^2$ for each $k$. Calculating $\|\mathbf{O}\|_{HS}$ requires $2n^2 - 1$ classical arithmetic steps. The Cholesky decomposition has various implementations but generally it requires $\frac{n^3}{3}$ classical arithmetic steps[2]. Thus $\frac{n^3}{3} + 2n^2 - 1$ classical steps should be added as an overhead cost (independent from the $k$ count) when evaluating the total cost of the quantum algorithm. In the meanwhile the classical complexity of evaluating $Tr(\mathbf{O}\rho(t))$ adds an overhead of $2n^3 - n^2$ to the original $4n^3 - 2n^2$ for each $k$.

Next we demonstrate the method proposed above on the same amplitude damping model used in the main text. To calculate the populations in the current basis $\{|0\rangle, |1\rangle\}$ we can construct $\mathbf{U}_{\mathcal{M}_k}$ and $\mathbf{U}_{\mathcal{N}_k}$ of the 2-dilation form in Eq. (8) using $\mathcal{M}_k = \mathbf{M}_k \otimes \mathbf{I}$, $\mathcal{N}_k = \mathbf{I} \otimes \bar{\mathbf{M}}_k$, and $\mathbf{D}_\mathbf{A} = \sqrt{\mathbf{I} - \mathbf{A}^\dagger \mathbf{A}}$ with $\mathbf{A} = \mathcal{M}_k, \mathcal{N}_k$:

$$\mathcal{M}_0 = \begin{pmatrix} 1 & 0 & 0 & 0 \\ 0 & 1 & 0 & 0 \\ 0 & 0 & \sqrt{e^{-\gamma t}} & 0 \\ 0 & 0 & 0 & \sqrt{e^{-\gamma t}} \end{pmatrix} \quad \mathcal{N}_0 = \begin{pmatrix} 1 & 0 & 0 & 0 \\ 0 & \sqrt{e^{-\gamma t}} & 0 & 0 \\ 0 & 0 & 1 & 0 \\ 0 & 0 & 0 & \sqrt{e^{-\gamma t}} \end{pmatrix}$$

$$\mathbf{D}_{\mathcal{M}_0} = \begin{pmatrix} 0 & 0 & 0 & 0 \\ 0 & 0 & 0 & 0 \\ 0 & 0 & \sqrt{1-e^{-\gamma t}} & 0 \\ 0 & 0 & 0 & \sqrt{1-e^{-\gamma t}} \end{pmatrix} \quad \mathbf{D}_{\mathcal{N}_0} = \begin{pmatrix} 0 & 0 & 0 & 0 \\ 0 & \sqrt{1-e^{-\gamma t}} & 0 & 0 \\ 0 & 0 & 0 & 0 \\ 0 & 0 & 0 & \sqrt{1-e^{-\gamma t}} \end{pmatrix} \quad (34)$$

$$\mathcal{M}_1 = \begin{pmatrix} 0 & 0 & \sqrt{1-e^{-\gamma t}} & 0 \\ 0 & 0 & 0 & \sqrt{1-e^{-\gamma t}} \\ 0 & 0 & 0 & 0 \\ 0 & 0 & 0 & 0 \end{pmatrix} \quad \mathcal{N}_1 = \begin{pmatrix} 0 & \sqrt{1-e^{-\gamma t}} & 0 & 0 \\ 0 & 0 & 0 & 0 \\ 0 & 0 & 0 & \sqrt{1-e^{-\gamma t}} \\ 0 & 0 & 0 & 0 \end{pmatrix}$$

$$\mathbf{D}_{\mathcal{M}_1} = \begin{pmatrix} 1 & 0 & 0 & 0 \\ 0 & 1 & 0 & 0 \\ 0 & 0 & \sqrt{e^{-\gamma t}} & 0 \\ 0 & 0 & 0 & \sqrt{e^{-\gamma t}} \end{pmatrix} \quad \mathbf{D}_{\mathcal{N}_1} = \begin{pmatrix} 1 & 0 & 0 & 0 \\ 0 & \sqrt{e^{-\gamma t}} & 0 & 0 \\ 0 & 0 & 1 & 0 \\ 0 & 0 & 0 & \sqrt{e^{-\gamma t}} \end{pmatrix} \quad (35)$$

With an initial $\rho = \frac{1}{4}\begin{pmatrix} 1 & 1 \\ 1 & 3 \end{pmatrix}$, $\|\rho\|_{HS} = \frac{\sqrt{3}}{2}$, the input state is:

$$\mathbf{v}_0 = \frac{1}{\|\rho\|_{HS}} \left( \mathbf{v}_\rho^T, \overbrace{0,...,0}^{m} \right)^T = \frac{1}{2\sqrt{3}} \left( 1,1,1,3,\overbrace{0,...,0}^{m} \right)^T \quad (36)$$

where $m = 8$ here is the number of zeros after $\mathbf{v}_\rho^T$. We use the same parameters as in the main text: set $\gamma = 1.52 \times 10^9 \text{s}^{-1}$ (typical nanosecond lifetime), numerically calculate $\mathbf{U}_{\mathcal{N}_k} \mathbf{U}_{\mathcal{M}_k} \mathbf{v}_0$ from $t = 0$ to $t = 1000$ ps with a time step of 10 picosecond, and obtain the populations of the ground and excited states from the first and fourth entries of $\mathbf{U}_{\mathcal{N}_k} \mathbf{U}_{\mathcal{M}_k} \mathbf{v}_0$. The results are the same as the smooth lines in Figure 1 in the main text. To calculate the populations in another basis $\{|+\rangle, |-\rangle\}$ where $|\pm\rangle = \frac{1}{\sqrt{2}}(|0\rangle \pm |1\rangle)$, we need the transformation matrix $\mathbf{T} = \frac{1}{\sqrt{2}}\begin{pmatrix} 1 & 1 \\ 1 & -1 \end{pmatrix}$. Now numerically calculate $\begin{pmatrix} \mathbf{T} \otimes \mathbf{I} & & \\ & \mathbf{I} & \\ & & \mathbf{I} \end{pmatrix} \begin{pmatrix} \mathbf{I} \otimes \overline{\mathbf{T}} & & \\ & \mathbf{I} & \\ & & \mathbf{I} \end{pmatrix} \mathbf{U}_{\mathcal{N}_k} \mathbf{U}_{\mathcal{M}_k} \mathbf{v}_0$ and we can obtain the populations of the $|\pm\rangle$ states. The results are the same as the smooth lines in Figure 2 in the main text. Now we evaluate the expectation value of an observable $\langle \mathbf{O} \rangle$ for $\mathbf{O} = \begin{pmatrix} -2 & 0.5 \\ 0.5 & 1 \end{pmatrix}$ the same as used in the main text. We construct $\mathbf{U}_{(\mathbf{I} \otimes \overline{\mathbf{L}}^\dagger)}$, $\mathbf{U}_{(\mathbf{L}^\dagger \otimes \mathbf{I})}$, $\mathbf{U}_{\mathcal{M}_k}$ and $\mathbf{U}_{\mathcal{N}_k}$ of the 4-dilation form in Eq. (31) and apply them to the initial state $\mathbf{v}_0$ in the form of Eq. (36) with $m = 16$. Numerically calculating the output vector will give us $\langle \mathbf{O} \rangle$ by Eq. (33). The results are the same as the smooth line in Figure 3 in the main text.

### 4. Quantum circuits used for the IBM Qiskit and Q 5 Tenerife device.

To implement Method 2 on the IBM simulator and quantum device, we further decompose the unitary gates into 1-qubit and 2-qubit elementary gates and construct the quantum circuits. All the gates used below are standard. For each circuit, only the $\theta$ parameter of the

$$U_3 = (\theta, \phi, \lambda) = \begin{pmatrix} \cos\frac{\theta}{2} & -e^{i\lambda}\sin\frac{\theta}{2} \\ e^{i\phi}\sin\frac{\theta}{2} & e^{i(\lambda+\phi)}\cos\frac{\theta}{2} \end{pmatrix}$$ gate changes during the time evolution. The circuits

below show the $U_3$ at the last time step at 991ps.

First for evolution in the original basis:

$\mathbf{U}_{\mathbf{M}_0} \mathbf{v}_1$:

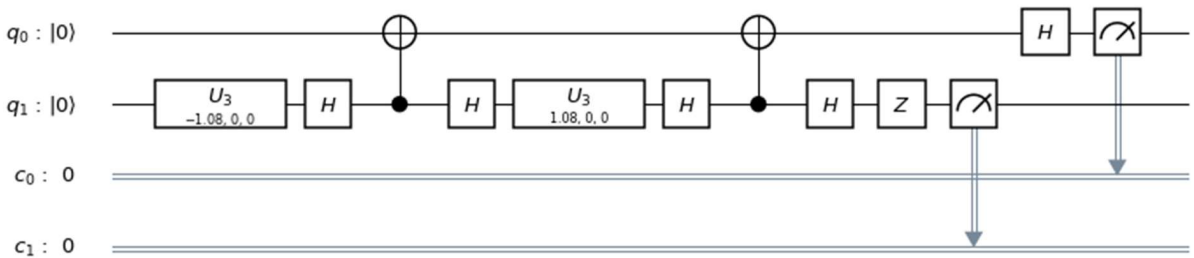

$\mathbf{U}_{\mathbf{M}_0} \mathbf{v}_2$:

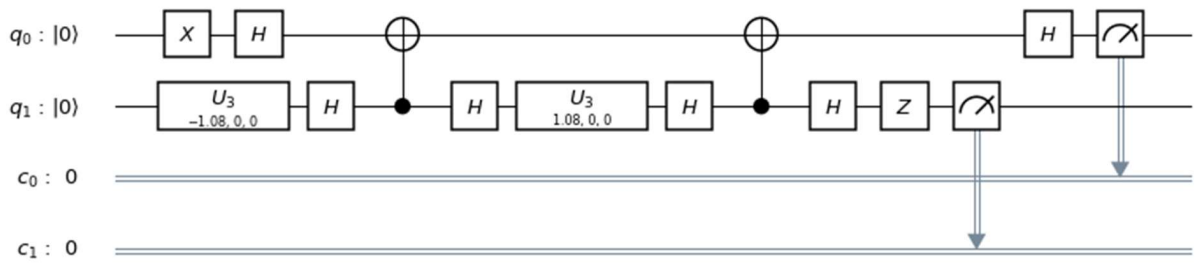

$\mathbf{U_{M_1} v_1}$:

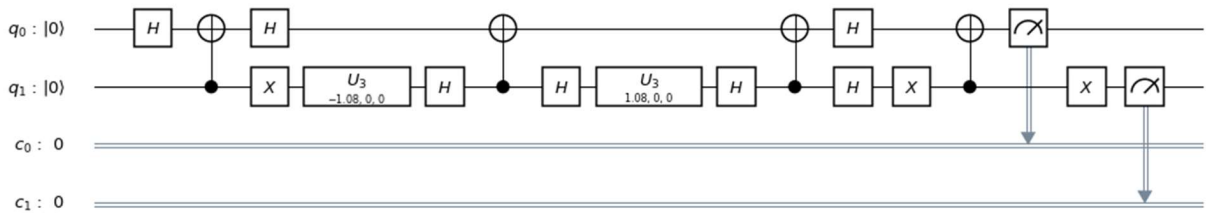

$\mathbf{U_{M_1} v_2}$:

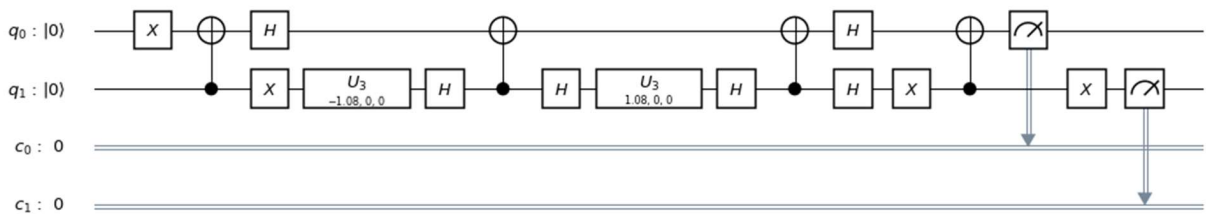

Next for evolution with a basis transformation:

$\begin{pmatrix} \mathbf{T} & 0 \\ 0 & \mathbf{I} \end{pmatrix} \mathbf{U_{M_0} v_1}$:

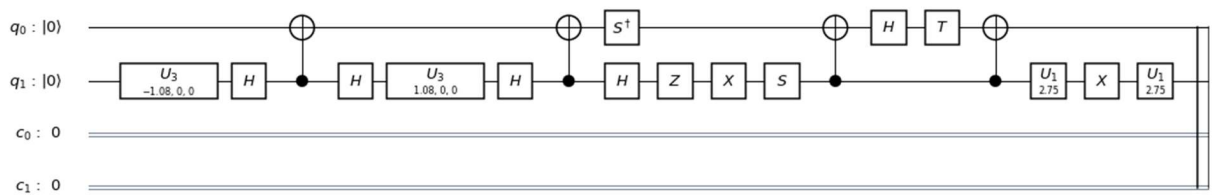
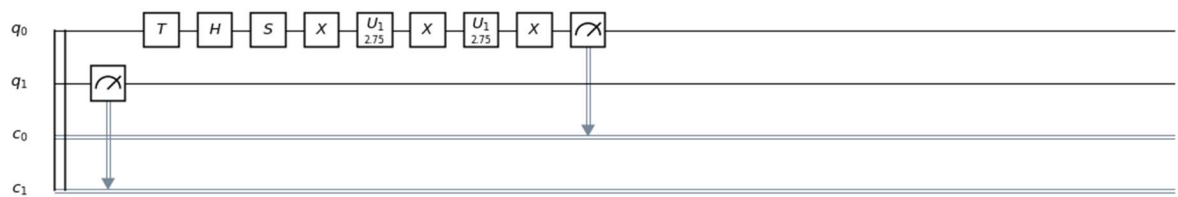

$$\begin{pmatrix} \mathbf{T} & 0 \\ 0 & \mathbf{I} \end{pmatrix} \mathbf{U}_{\mathbf{M}_0} \mathbf{v}_2 :$$

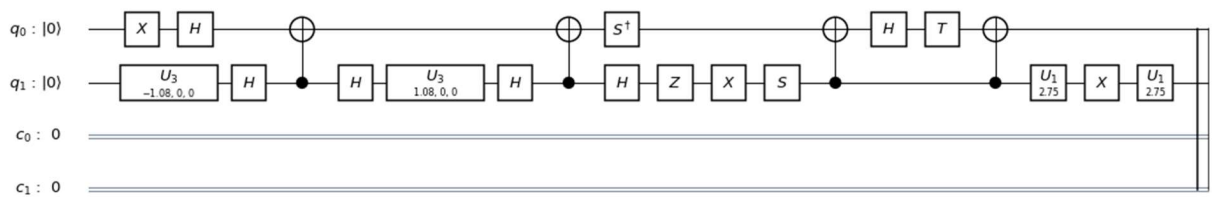
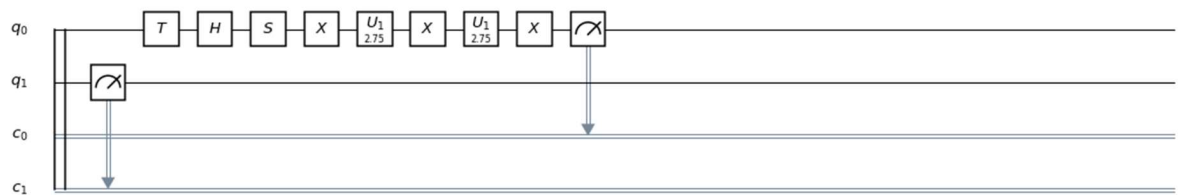

$$\begin{pmatrix} \mathbf{T} & 0 \\ 0 & \mathbf{I} \end{pmatrix} \mathbf{U}_{\mathbf{M}_1} \mathbf{v}_1 :$$

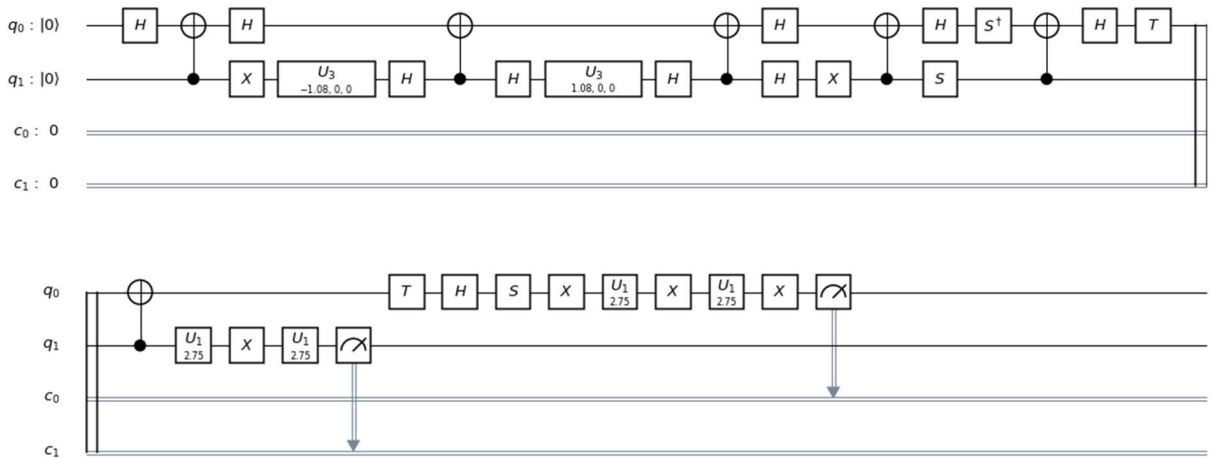

$$\begin{pmatrix} T & 0 \\ 0 & I \end{pmatrix} U_{M_1} v_2:$$

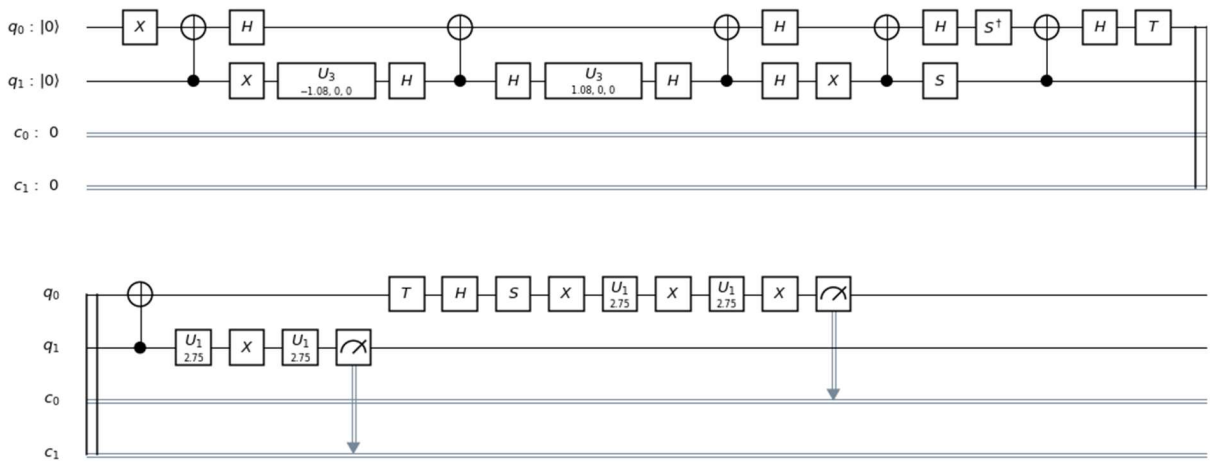

For the evolution with $\langle \tilde{O} \rangle$ evaluation, due to the large number of gates involved we only show the circuit for $U_{L^\dagger} U_{M_0} v_1$:

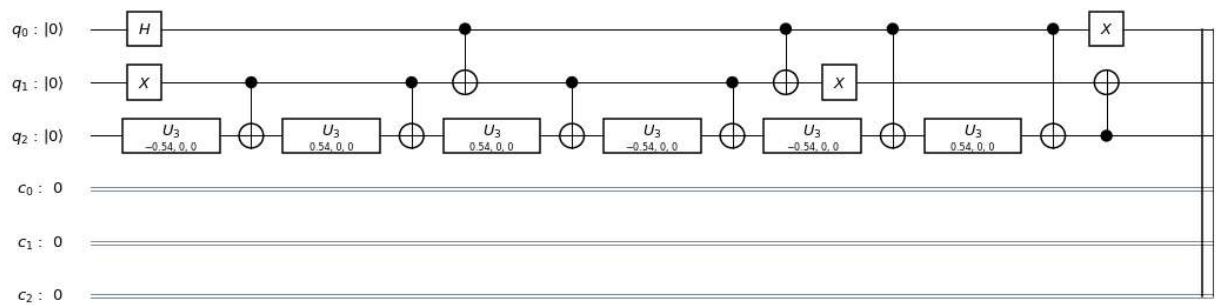
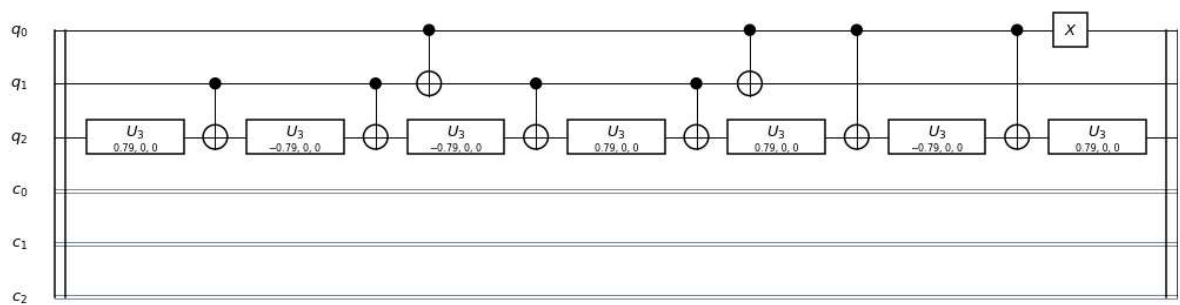
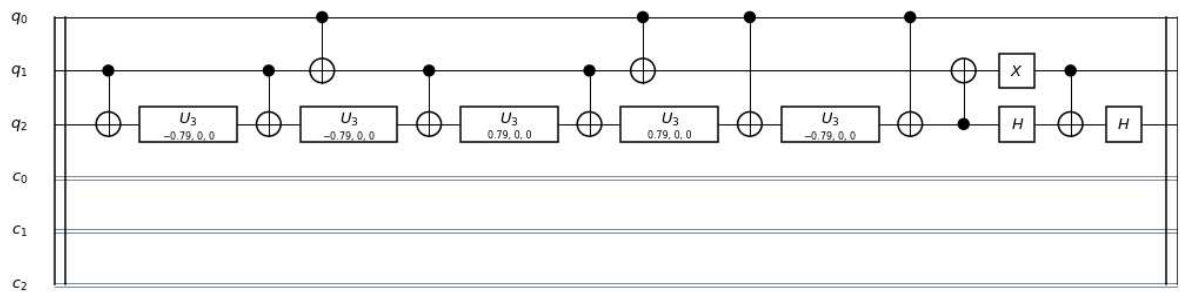
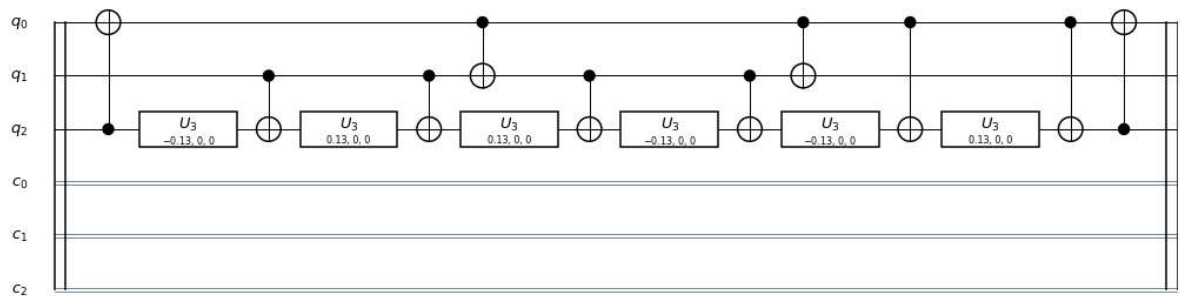

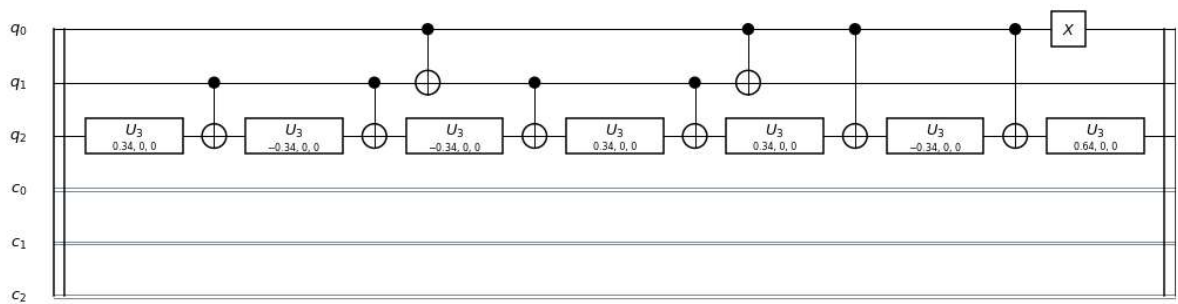

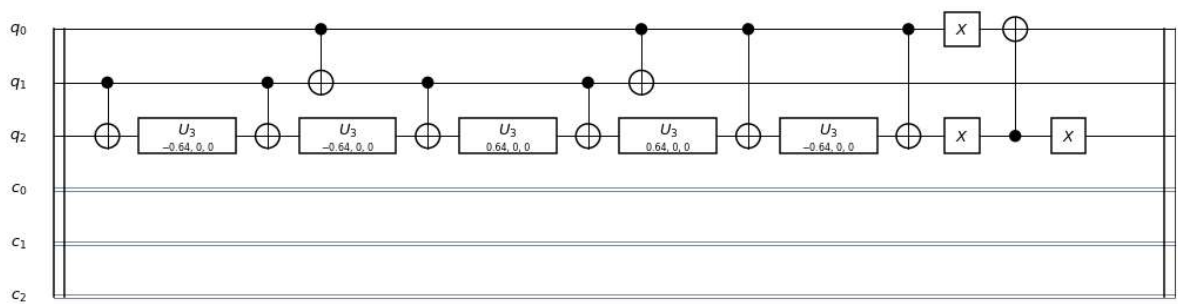

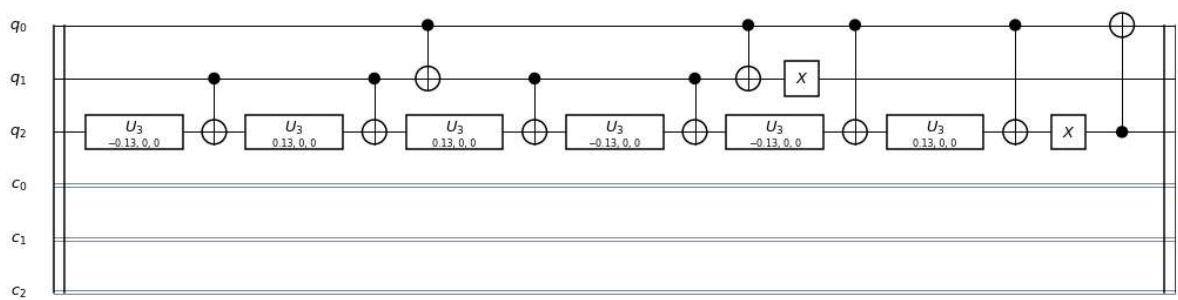

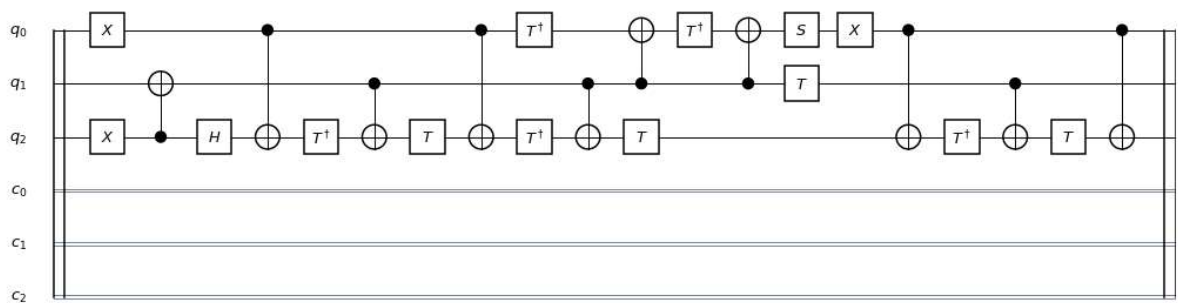

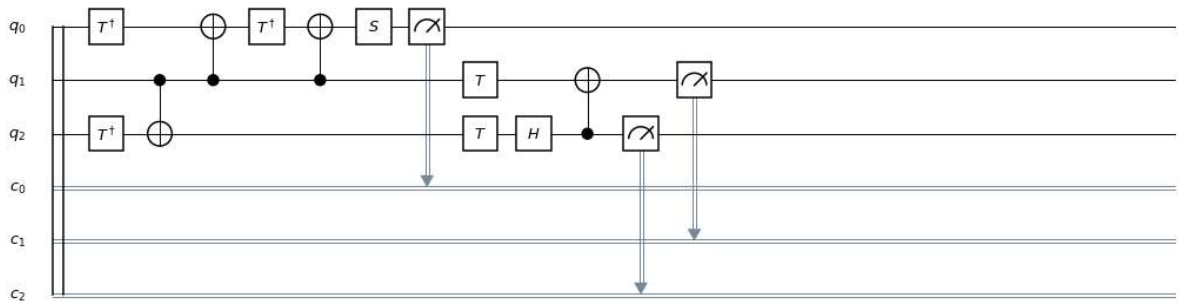